# Reproducible noise in a macroscopic system: magnetic avalanches in Perminvar

J. S. Urbach*, R. C. Madison, and J. T. Markert

*Center for Nonlinear Dynamics and Department of Physics,
The University of Texas at Austin, Austin, TX 78712*

(August 16, 1995)

A study of magnetic avalanches in Perminvar, an Fe-Ni-Co alloy, shows that some avalanches are almost exactly reproducible from one magnetic field cycle to the next, while others show significant variability. Averaging over many cycles produces a fingerprint reflecting the reproducibility of the noise. The fingerprint is not strongly temperature or driving-rate dependent, indicating that the variability is a consequence of dynamical effects. We also find that the slope of the cycle-averaged magnetization, $d\langle M\rangle/dH$, is correlated with the cycle-to-cycle variations in magnetization, $\langle(\delta M)^2\rangle$.

When a slowly varying magnetic field is applied to a ferromagnet, the response is usually dominated by a sequence of abrupt jumps, or avalanches, as the system moves from one metastable state to another. This characteristic, known as Barkhausen noise, is the origin of magnetic hysteresis [1,2]. Barkhausen noise is usually modeled with simple magnetic dynamics interacting with quenched disorder, and the avalanche sequence is a result of the interaction between the two [3]. At zero temperature, the avalanches are exactly reproducible: if the system is repeatedly prepared in a particular configuration, the sequence of avalanches observed as the field is changed will be identical on each repetition. A system exhibiting the return-point memory effect, such as the random-field Ising model [4], will always return to its initial configuration after a field cycle, so the same sequence of avalanches is produced on each cycle.

In experiments on macroscopic magnets, by contrast, the only correlation between the avalanche sequence on subsequent cycles is the average rate at which avalanches occur, which determines the shape of the hysteresis loop. The absence of reproducibility is a natural result of the combination of environmental noise, including thermal effects, and the dynamic complexity of systems with many interacting degrees of freedom. As a consequence, there has been very little investigation of the reproducibility of magnetic noise [5].

We have been studying the Barkhausen effect in an unusual Fe-Ni-Co alloy that allows us to repeatedly prepare the magnet in a particular configuration [6]. We have found that the avalanche sequence observed as the system is forced out of this configuration has features that are almost exactly reproducible (events of a particular size that occur at the same value of applied field on each cycle) intermingled with events that show no apparent reproducibility (Fig. 1). The behavior is frequency independent (at low driving rates) and at most weakly temperature dependent, suggesting that dynamic instability plays an important role in the observed behavior.

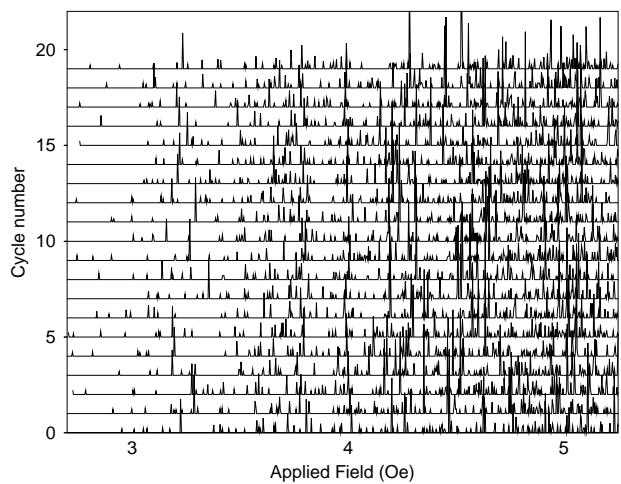

FIG. 1. Magnetic avalanches in Perminvar on successive magnetic field cycles. Data from the increasing field portion of a -6 to 6 Oe triangle wave applied field (0.32 Oe/sec). Each line represents voltage vs. field for one cycle, offset by 50 $\mu V$ to show cycle to cycle variability.

The reproducibility is less evident at higher fields, after a large number of avalanches have occurred on each cycle. The avalanche activity averaged over many field cycles, however, shows clear fluctuations as a function of applied field. Thus the averaging produces a fingerprint, analogous to effects observed in mesoscopic samples, that is presumably a reflection of the quenched disorder in the material. Finally, we have discovered a remarkable correlation between this averaged activity and the cycle-to-cycle variation in the magnetization. This result suggests an analogy with the fluctuation-dissipation relationship of equilibrium thermodynamics, with cycle averaging replacing thermal averaging.

The magnetization changes of a ferromagnet are dominated by the motion of magnetic domain walls. The application of a magnetic field produces a force on the walls, causing domains that are aligned along the field to grow at the expense of domains which are magnetized in the opposite direction. This growth is very uneven and the wall configurations are usually far from equilibrium due to the pinning of the walls by inhomogeneities. When the applied field is changed very slowly, the changes in domain wall configuration are well separated, and the magnetization change can be viewed as a sequence of distinct jumps. These jumps are easily detected as voltage spikes in a pickup coil surrounding the sample.

Certain alloys of iron, nickel, and cobalt can be prepared in a state where a particular disordered domain wall configuration is strongly pinned. During the process of magnetic annealing, the easy axis (the preferred axis for magnetization) adjusts to lie along the direction of the actual magnetic field during the anneal [7]. The annealing temperature ($425-475°C$ for Perminvar) must be high enough to allow for atomic diffusion and low enough so that thermal fluctuations do not swamp the interaction between the directional order and the magnetic field. This process is often exploited by annealing a material in an applied field, resulting in a uniaxial sample and a square hysteresis loop (the magnetization of the entire sample reverses direction at the same field). In the absence of an applied field, the direction of the magnetic domains, and therefore the local field, will vary throughout the sample. The annealing process then results in a pattern of preferred directions that matches the configuration of domains present in the material at the time of the annealing. When the sample is cooled, the atomic arrangements are essentially fixed and the domain walls are strongly pinned. Consequently, when a magnetic field is applied, the magnetization change is initially limited to small rotations of the magnetization within the domains, which is an elastic process involving no hysteresis and no avalanches. At larger fields [8] the domain walls begin to move and hysteresis sets in. When the field is reduced to zero, the domain walls return exactly to their original positions, as evidenced by the absence of remanent magnetization. This is a result of the fact that the annealing process assures that the original configuration has a much lower energy than any other configuration. Once a sufficiently large field is applied, the original domain wall configuration is destroyed and the system never returns to its original state. At this point the reversible behavior can only be recovered by repeating the annealing sequence.

Figure 1 shows the Barkhausen signal obtained from a 5mm x 2mm x 0.1mm sample of Perminvar ($Fe_{30}Ni_{45}Co_{25}$) at room temperature, as the domain walls are being forced out of the strongly pinned configuration. There are no detectable avalanches at fields below about 1.5 Oe. (Details of sample preparation and experimental setup have been described previously [9].) The voltage from the pickup coil is proportional to $dM/dt$ and therefore $dM/dH$, since the field is increased linearly in time. The height of the pulse is a measure of the maximum domain wall velocity during the event, and the area underneath the pulse is proportional to the net change in magnetization [10]. The large peaks in Fig. 1 represent a magnetization change equal to approximately 15 $\mu m^3$ of magnetic moments switching directions.

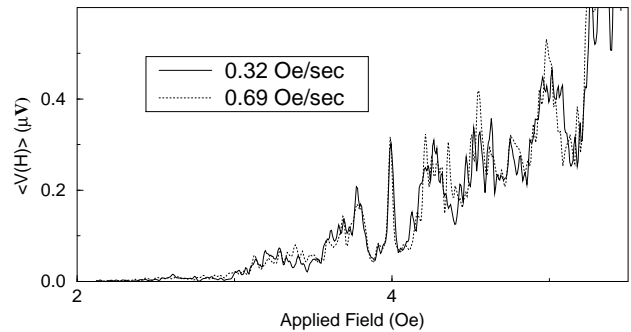

FIG. 2. Voltage due to the Barkhausen pulses, averaged over 75 magnetic field cycles, for two different driving rates.

The most notable feature in this sample is the peak at 4 Oe, which occurs with the same magnitude at almost exactly the same field on every cycle. Other peaks in the data seem to appear in every cycle, but with more variability in position and magnitude. Similar behavior has been observed in several samples. The degree of reproducibility is not well correlated with the size of the avalanches. For example, the avalanche around 3.8 Oe is usually bigger than the 3 Oe avalanche, but occurs over a wider range of fields. However, we have never observed significant reproducibility in the smallest avalanches.

We have investigated quantitative measures of the reproducibility by looking at the behavior of the Barkhausen signal as a function of field, averaged over many cycles [11]. If the signal was uncorrelated from one cycle to the next, this should result in a smooth curve, whereas a perfectly reproducible avalanche sequence would result in sharp spikes. The experimental result, shown in Fig. 2, lies somewhere in between. Each curve represents the average of 75 cycles, taken at two different driving rates. This fingerprint demonstrates that some reproducibility persists even after a cycle to cycle correlation can no longer be detected by direct comparison. The two curves in Fig. 2 are indistinguishable within the statistical noise, yet the driving rates differ by a factor of 2. This is consistent with the idea that the avalanches represent transitions between metastable states, and that we are in the low driving rate limit.

Similar reproducibility of noise or fluctuations has been seen in a variety of mesoscopic systems, most commonly in the field dependence of the conductivity [12]. In both cases, the fingerprint is a reflection of the quenched disorder. The annealing procedure described above freezes most of the domain walls in the system. The wall that is most weakly pinned will presumably be the first to move on each cycle. Since the wall is being moved a very small distance and only interacts with the disorder locally, our macroscopic system is in effect mesoscopic.

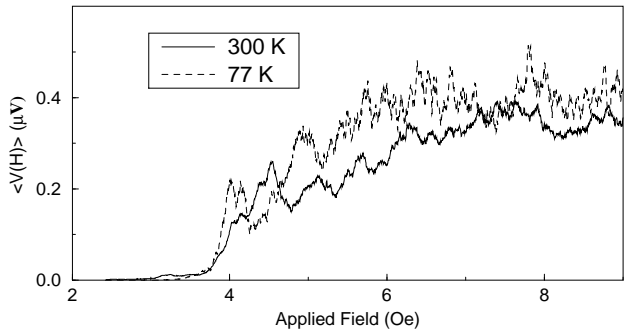

FIG. 3. Average voltage at room temperature and at 77 K (200 cycles each, 2.2 Oe/sec). The fluctuations below about 6 Oe are reproducible.

In order to understand the role of thermal fluctuations, we have looked at the temperature dependence of avalanche sequences. Unfortunately, the features move as the temperature changes, presumably due to the effects of thermal expansion of the sample (magnetostriction). We have found, however, that the cycle-to-cycle variability is not a strong function of temperature. Figure 3 shows the cycle averaged noise signal for a different sample, at 77 K and 300 K. This sample lacks the dramatic peak of the sample from Figs. 1-2, but the fluctuations observable in the average are well above the statistical noise. Despite a change in temperature of a factor of 4, the size of the variations in the average signal change by less than 50%. Other measures of temperature dependence are consistent with this semi-quantitative conclusion [13].

Our results appear to be inconsistent with a model of domain wall dynamics that considers a rigid wall interacting with quenched disorder [14]. Since the only degree of freedom in this model is the position of the wall along a line, the system necessarily travels through the same series of metastable states. The effect of a finite temperature (or other additive stochastic process) will be to allow the system to escape each state a little earlier than it would otherwise. The amount of the shift will be strongly temperature dependent, and should also depend on the driving rate. The cycle-to-cycle variability will depend on the relevant time scales in the problem, but should be strongly temperature and rate dependent as well.

If more complex domain wall dynamics are considered, all of the observed behavior can be accounted for. As a domain wall moves across the sample, it may encounter the same disorder on each trip, but the configuration of a flexible wall can vary from one cycle to the next. Since the force (applied field) necessary to move the wall depends on the configuration of the wall, there will be some variability in each avalanche. The divergence of two neighboring trajectories (e.g. successive movements of the domain wall, with identical initial conditions but different realizations of environmental noise) in deterministic systems is a now familiar feature in nonlinear dynamics. In such a situation the cycle-to-cycle variability could be almost independent of the strength of the noise and the driving rate. An arbitrarily small amount of noise, coupled with the dynamics, will allow the system to sample all possible routes through the disorder, and the variability is a measure of the distribution of possible paths of the domain wall. The presence of quenched disorder limits the extent to which trajectories can diverge, and is essential for the observed reproducibility. A strong pinning center could effectively funnel all possible trajectories into a particular metastable state, resulting in a reproducible avalanche. This picture assumes that there are no changes in the domain wall topology [15]. We are not aware of any theoretical investigations of the interplay between stochastic noise, nonlinear dynamics, and quenched disorder that address these questions.

We have discovered a very intriguing correlation that suggests an analogy with equilibrium thermodynamics. We can determine the magnetization according to

$$M_i(H) = \int_0^H V_i(H')dH'$$

where $V_i(H)$ is the voltage from the pickup coil on the $i^{th}$ magnetic field cycle. The cycle-averaged magnetization is given by

$$\langle M(H) \rangle = \frac{1}{N} \sum_{cycles} M_i(H)$$

where N is the number of magnetic field cycles included in the average. This value of M is necessarily less than the true magnetization of the sample because it does not include the reversible magnetization changes or the avalanches that are smaller than our threshold. $\langle V(H) \rangle$ shown in Figs. 2 and 3 is equal to $d\langle M \rangle/dH$, the nonequilibrium susceptibility of the sample. (When the field direction is reversed, no avalanches are observed initially, so for decreasing field, $d\langle M \rangle/dH = 0$). According to the fluctuation-dissipation theorem, the susceptibility is proportional to the size of the magnetization fluctuations in equilibrium [16]. Figure 4 shows the fluctuations in the cycle-averaged magnetization, $\langle (\delta M)^2 \rangle = \langle M^2 \rangle - \langle M \rangle^2$ on the same plot as the nonequilibrium susceptibility. $\langle (\delta M)^2 \rangle$ is scaled by an arbitrary factor, of order $(\Delta M)(\Delta H)$, where $\Delta M$ is the size of the large avalanches, and $\Delta H$ is the field between avalanches. A strong correlation between $d\langle M \rangle/dH$ and $\langle (\delta M)^2 \rangle$ has been seen in all of our samples, although the very sharp features such as the peak at 4 Oe in Fig. 2 are much less dramatic in $\langle (\delta M)^2 \rangle$. It perhaps not surprising that the cyclic variation of the magnetization increases at those fields where the magnetization is changing rapidly. If we imagine the system as moving from one strongly pinned configuration to another, with some cyclic variability in the depinning field, the fluctuations will be large when the transitions are made, and low when they gather at the pinning sites. The demagnetization field, which creates an equilibrium position for the domain wall [9] may also play a role in limiting the divergence of different trajectories. Nonetheless, the extent to which the two quantities track each other is quite remarkable, particularly when the fluctuations in $\langle(\delta M)\rangle^2$ and $d\langle M\rangle/dH$ are small compared to their total value. As we have argued above, our results suggest that the cyclic variations of the field coupled with the domain wall dynamics allow the system to explore different configurations. This produces a certain amount of ergodicity, and perhaps some nontrivial similarity to the equilibrium system.

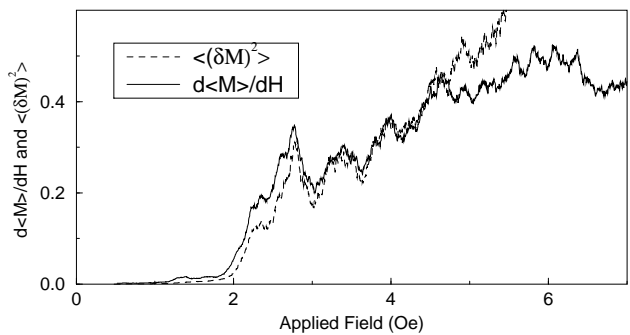

FIG. 4. $\langle V(H)\rangle$ (or $d\langle M\rangle/dH$) and $\langle(\delta M)^2\rangle$ averaged over 400 cycles. (2.2 Oe/sec, 300 K; similar behavior is observed at 77 K.) The units of the abcissa are $\mu V$, and $\langle(\delta M)^2\rangle$ is scaled by an arbitrary factor (see text).

We would like to thank Michael Marder and Bala Sundaram for helpful discussions. This work was supported by the DOE, Office of Basic Energy Sciences, the Robert A. Welch Foundation under Grant No. F1191, and the NSF under Grant No. DMR-9158089.